\newcommand{\be}{\begin{equation}}
\newcommand{\ee}{\end{equation}}
\newcommand{\ben}{\begin{eqnarray}}
\newcommand{\een}{\end{eqnarray}}
\newcommand{\bc}{\begin{center}}
\newcommand{\ec}{\end{center}}

\documentstyle{lamuphys}
\makeatletter
\let\chapter\hid@chapter
\makeatother

\begin{document}
\pagenumbering{arabic}

\title{\bf Boson stars in the centre of galaxies?}

\author{Franz E.~Schunck and Andrew R.~Liddle}

\institute{Astronomy Centre, University of Sussex, Falmer, Brighton BN1 9QJ,
UK\\
emails: fs@astr.maps.susx.ac.uk, a.liddle@sussex.ac.uk}

\maketitle

\begin{abstract}
We investigate the possible gravitational redshift values for boson
stars with a self-interaction, studying a wide range of possible
masses. We find a limiting value of $z_{\rm lim} \simeq 0.687$ for
stable boson star configurations. We can exclude the
direct observation of boson stars. X-ray spectroscopy is perhaps the most
interesting possibility.
\end{abstract}

\section{Introduction}

The idea of the boson star goes back to Kaup (\cite{kau}). A
boson star is a gravitationally bound collection of bosonic particles,
arising as a solution of the Klein--Gordon equation coupled to general
relativity. Many investigations of the possible configurations have
been carried out; for reviews see \cite{rev,jet,arl}. For
non-self-interacting bosons of mass $m$, the mass of a typical
configuration is of order $m^2_{{\rm Pl}}/m$, to be compared with a
typical neutron star mass of $m^3_{{\rm Pl}}/m^2_{{\rm neutron}}$
which is about a solar mass. Here $m_{{\rm Pl}}$ is the Planck
mass.

The situation is very different if the boson stars have even a very
weak self-interaction. \cite{col} showed
that the maximum mass of stable configurations is then of order
$\lambda^{1/2} m^3_{{\rm Pl}}/m^2$, where $\lambda$ is the scalar
field self-coupling, normally assumed to be of order unity. Then boson
star configurations exist with mass (and radius) similar to that of
neutron stars, if the bosons, like neutrons, have a mass around 1
GeV. They can also be much heavier, should the bosons be lighter.
We allow ourselves to consider a very wide range of
possibilities for the boson star mass and radius. If boson stars
exist, they provide an alternative explanation for stellar systems
in which an object is inferred to have a high mass; conventionally, a
`star' with mass greater than a few solar masses is assumed to be a
black hole.

We investigate
the implications of assuming that the material from which boson stars
are made interacts with neighbouring baryonic material and photons just
gravitationally, as the relation between a visible galaxy and its
dark matter halo. An example already existing in the
literature is the {\em boson--fermion star} (\cite{bosfer,bosfer2}),
which is made up of bosons and neutrons interacting only 
gravitationally. However, while a galaxy halo can be described using 
Newtonian theory, boson stars close to the 
maximum allowed mass are general relativistic objects. This gives a new 
characteristic of such objects, a gravitational redshift (\cite{lid}).

The boson star model is described by
the Lagrange density of a massive complex self-gravitating scalar
field
$
{\cal L} = \sqrt{\mid g \mid} \left[
  m_{{\rm Pl}}^2 R/8\pi+ \partial_\mu \Phi^\ast \partial^\mu \Phi 
- U(|\Phi |^2) \right ]/2
$,
where $R$ is the curvature scalar, $g$ the determinant of the metric
$g_{\mu \nu }$, and $\Phi$ is a {\em complex} scalar field with a
potential $U$. We take $\hbar=c=1$. We want to
have an additional global $U(1)$ symmetry (conserved particle number),
so we can take the following potential
$
U = m^2 |\Phi |^2 + \lambda |\Phi |^4 /2
$,
where $m$ is the scalar mass and $\lambda$ a dimensionless constant
measuring the self-interaction strength. For
spherically symmetric solutions we use the static line element
$
ds^2 = \exp (\nu (r)) dt^2 - \exp(\mu (r)) dr^2
  - r^2 ( d\vartheta^2 + \sin^2\vartheta \, d\varphi^2)
$.
The most general scalar field ansatz consistent with this metric is
$
\Phi (r,t) = P(r) \exp (-i \omega t)
$,
where $\omega$ is the frequency.

\section{Gravitational redshift and detectability of boson stars}

The maximum possible redshift for a given
configuration is obtained if the emitter is exactly at the center $R_{{\rm
int}}=0$. The receiver is always practically at infinity. For all 
other redshifts in between, we define the redshift function
$
1+z_{{\rm g}}(x) \equiv \exp \left(-\nu (x)/2\right)
$,
where $x=\omega r$. A boson star with the maximum mass gives the
highest value one
can obtain from stable configurations (unstable configurations can
yield very high redshift values) (\cite{kus}). We find
that with increasing self-interaction values $\Lambda := \lambda
m_{{\rm Pl}}^2/4\pi m^2$
also the maximal redshift value grows (\cite{lid}).
For $\Lambda \rightarrow \infty $, we find
the asymptotic value $z_{{\rm lim}} \simeq 0.687$.

The mass $M$ of a boson star composed of non-self-interacting
particles is inversely proportional to $m$, while the mass of a
self-interacting boson star is proportional to $\sqrt{\lambda }/m^2$;
see \cite{col}. Taking $\lambda \sim 1$, then for small $m$ (to be
precise, provided $\Lambda \gg 1$) the self-interacting star is much
more massive.  For example, if we want to get a boson star with a mass
of order $10^{33}$g (a solar mass), then we need $m \sim 10^{-10}$ eV
for $\lambda = 0$, or $m \propto \lambda^{1/4}$ GeV if $\lambda \gg
10^{-38}$ (we see that the self-coupling has to be extraordinarily
tiny to be negligible). In this example, the scalar particle has a
mass comparable to a neutron, leading to a boson star with the
dimensions of a neutron star. If we reduce the scalar mass further, to
$m \sim 1$ MeV, then we find $M \sim 10^{39} \sqrt{\lambda }$ g and
$R\sim 10^6 \sqrt{\lambda }$ km; this radius is comparable to that of
the sun, but encloses 10$^6$ solar masses. These parameters are
reminiscent of supermassive black holes, for example as in Active
Galactic Nuclei; the mass--radius relation is effectively fixed just
by the objects being relativistic.
In all cases, the density of the boson stars makes their direct
detection as difficult as in the case of black holes; in particular,
they cannot be resolved in any waveband, cf.~(\cite{lid}).

However, even if boson stars cannot be directly resolved, their
influence might still be visible if material in their vicinity is
sufficiently luminous. It is necessary to find a certain amount of
luminous matter within the gravitational potential of the boson star.
This could, for example, be H{\small I} gas clouds as seen in
galaxies. One might also expect accretion discs about boson
stars, though there the luminosity could be dominated by regions
outside the gravitational potential and the boson star would be
indistinguishable from a black hole.

The most promising technique for observing supermassive boson stars is
to consider a wave-band where they might be extremely luminous, e.g.~X-rays.
A very massive boson star,
say $10^6 M_{\odot}$ is likely to form an accretion disk, and since its
exterior solution is Schwarzschild it is likely to look very similar
to an AGN with a black hole at the center. In X-rays, it has been
claimed by \cite{Ietal} that using ASCA data they have
probed to within 1.5 Schwarzschild radii. A boson star
configuration provides a non-singular solution where
emission can occur from arbitrarily close to the center. The
signature they use is a redshifted wing of the Iron K-line. If such
techniques have their validity confirmed, it may ultimately be
possible to use X-ray spectroscopy to map out the shape of the
gravitational potential close to the event horizon or boson
star.

The rotation curves about a boson star (\cite{lid}) show an increase up
to a maximum with more than one-third of the velocity of light followed
by a Keplerian decrease. If boson stars exist,
then such enormous rotation velocities are not necessarily signatures
of black holes. In \cite{sch2}, a model with massless bosonic
particles was applied to fit rotation curve data of spiral and dwarf galaxies.

\section*{Acknowledgments}
FES is supported by a Marie Curie research fellowship (European Union
TMR programme) and ARL is supported by the Royal Society. We would
like to thank John Barrow and Eckehard W.~Mielke for helpful discussions
and comments.


\nonfrenchspacing

\end{document}